\documentclass[prl,twocolumn,superscriptaddress,showpacs,floatfix]{revtex4}
\usepackage{graphicx}% Include figure files
\usepackage{dcolumn}% Align table columns on decimal point
\usepackage{bm}% bold math

\begin{document}

\title{Ground-state and single-particle energies of nuclei around
$^{16}$O, $^{40}$Ca, and $^{56}$Ni
from realistic nucleon-nucleon forces}

% Force line breaks with \\

\author{S. Fujii}
 \email{sfujii@rche.kyushu-u.ac.jp}
\affiliation{
Center for Research and Advancement in Higher Education,
Kyushu University,
Fukuoka 819-0395, Japan
}

\author{R. Okamoto}
\affiliation{
Department of Physics,
Kyushu Institute of Technology,
Kitakyushu 804-8550, Japan
}

\author{K. Suzuki}
\affiliation{
Department of Physics,
Kyushu Institute of Technology,
Kitakyushu 804-8550, Japan
}

\date{\today}% It is always \today, today,

\begin{abstract}
We perform {\it ab initio} calculations for nuclei around $^{16}$O,
$^{40}$Ca, and $^{56}$Ni using realistic nucleon-nucleon forces.
In particular, $^{56}$Ni is computed as the heaviest nucleus
in this kind of {\it ab initio} calculation.
Ground-state and single-particle energies including three-body-cluster
effects are obtained within the framework of the unitary-model-operator
approach.
It is shown that the CD-Bonn nucleon-nucleon potential gives
quite good results close to the experimental values for all nuclei
in the present work.

\end{abstract}


\maketitle

One of the most fundamental problems in nuclear physics is to
describe and understand nuclear properties from the underlying
nuclear forces.
To solve this problem, a realistic nucleon-nucleon ({\it NN})
interaction~\cite{Machleidt89}, which has a strong repulsive core and
a complicated spin-isospin structure,
has been employed~\cite{Kamada01}.
In addition, to obtain more quantitative results,
a three-nucleon ({\it NNN}) interaction~\cite{Fujita57, Pieper01}
has been used in some cases.
Light nuclei with the mass number up to $A\simeq12$ have been
well understood from the {\it ab initio} calculations employing
the realistic {\it NN} and {\it NNN} interactions
with the Green's function Monte Carlo (GFMC)
method~\cite{Pieper01,Pieper01-2}
and the no-core shell model (NCSM)~\cite{Navratil00,Navratil00-2}.
While these methods have been successful in the light nuclei,
their applications to heavier systems become rather difficult due to
exponential increase of the computer performance to be needed.

Coupled-cluster (CC) theory or, in other words, $e^{S}$ (or $e^{T}$)
method~\cite{Coester58, Bartlett07} is a promising one for
the microscopic calculation for the heavier nuclei.
Recently, the first calculations for nuclei up to
the $pf$-shell region, $^{48}$Ca and $^{48}$Ni,
have been reported with a CC method including the excitations of
the singles and doubles which is referred to as CCSD~\cite{Hagen08}.
Well-converged results in a sufficiently large model space have been
obtained using a chiral N$^{3}$LO {\it NN} interaction~\cite{Entem03}
as one of the realistic {\it NN} forces.
While the CCSD calculations give results fairly close to
the experimental values, there still remain some discrepancies
between the results and experiments.
One of the reasons of the discrepancies may be attributed to
the missing {\it NNN} interaction in the calculation.
Although the {\it NNN} force has been considered to be an indispensable
ingredient for a more quantitative description of the nuclear
properties, there has been no definite way of using the {\it NNN} force
directly in the calculation for the heavier nuclei.

Given this situation, it is still worthy to compute
nuclear properties using only the realistic {\it NN} interaction
in a rigorous way and to investigate to what extent nuclei can be
described with only the {\it NN} force.
Such a study could be helpful to evaluate the magnitude
of the {\it NNN}-force effect in heavier nuclei in future works.

In this Letter, we report the results of calculated ground-state
energies and single-particle ones for hole states in nuclei around
$^{16}$O, $^{40}$Ca, and $^{56}$Ni with the unitary-model-operator
approach (UMOA)~\cite{Suzuki87, Suzuki94, Fujii04}.
The calculation for $^{56}$Ni, which is a typical $pf$-shell nucleus,
is performed for the first time within the UMOA framework.
Furthermore, $^{56}$Ni is the heaviest nucleus for which this kind of
{\it ab initio} calculation has been performed.
In the UMOA, a Hermitian effective interaction is derived
through a unitary transformation~\cite{Okubo54, Suzuki82}.
The unitary transformation method has been widely used
in other microscopic methods in nuclear physics,
such as the NCSM~\cite{Navratil00,Navratil00-2}
and the hyperspherical harmonics effective interaction method
(EIHH)~\cite{Barnea00}.
The unitary transformation treats successfully short-range correlations
due to the strong repulsive core of the {\it NN} force in
a {\it truncated} model space, but the model space should be
sufficiently large in the sense of the {\it ab initio} calculation.

In the UMOA, a unitarily transformed Hamiltonian $\tilde{H}$ of
the original many-body Hamiltonian $H$ is given in a cluster-expansion
form as
$\tilde{H}=e^{-S}He^{S}=\tilde{H}^{(1)}+\tilde{H}^{(2)}+\tilde{H}^{(3)}
+\cdot \cdot \cdot$, where $S$ is a two-body anti-Hermitian operator
and is determined by solving a decoupling equation between
the model space and its compliment~\cite{Suzuki80}.
The $\tilde{H}^{(1)}$, $\tilde{H}^{(2)}$, and $\tilde{H}^{(3)}$
are the one-, two-, and three-body cluster (3BC) terms, respectively.
The method of the actual calculation and the results of nuclei around
$^{16}$O including up to the two-body cluster terms using modern {\it NN}
forces have been given in detail in our previous study~\cite{Fujii04}.
In the present work, we apply this method to the heavier nuclei up to $^{56}$Ni
and evaluate effects of the 3BC terms systematically.
As for the evaluation of the 3BC terms, we follow the prescription
given in Refs.~\cite{Suzuki87, Suzuki94}.

\begin{figure}[t]
\includegraphics[width=.360\textheight]{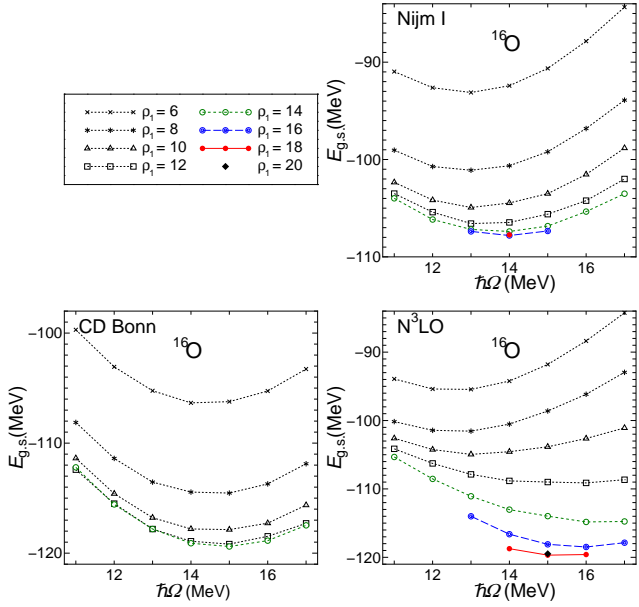}
\caption{\label{fig:o16egs} (Color online) The $\hbar \Omega$ and
$\rho_{1}$ dependences of the calculated ground-state energies
$E_{\rm g.s.}$ of $^{16}$O.}
\end{figure}

In Fig.~\ref{fig:o16egs}, we first demonstrate $\hbar \Omega$ and
$\rho_{1}$ dependences of the calculated ground-state energies of
$^{16}$O including the 3BC effects.
Here, $\hbar \Omega$ is the harmonic-oscillator (h.o.) energy of
the single-particle basis states.
The $\rho_{1}$ stands for a boundary number defined with a set of
the h.o. quantum numbers $\{n_{1},l_{1}\}$ and $\{n_{2},l_{2}\}$
of the two-body states as $\rho_{1}=2n_{1}+l_{1}+2n_{2}+l_{2}$,
and specifies the size of the model space of the two-body states.
The Nijm-I~\cite{Stoks94}, the CD-Bonn~\cite{Machleidt96},
and the chiral N$^{3}$LO~\cite{Entem03} {\it NN} forces
are employed as the realistic {\it NN} interactions.
The Coulomb force is added to the proton-proton channel.
It is seen that well-converged results with respect to the boundary
number $\rho _{1}$ are obtained at $\rho_{1}=18$ and $14$ for
the Nijm-I and the CD-Bonn interactions, respectively.
For the N$^{3}$LO interaction, the convergence property is rather
different from the other two forces.
For example, at $\hbar \Omega =15$ MeV, it is difficult to search for
convergence up to $\rho _{1}=14$.
However, for the larger values of $\rho_{1}$, the results rapidly
converge toward the point at $\rho _{1}=20$.

\begin{table}[b]
\caption{\label{tab:o16_egs} The calculated energies of
the one- and two-body cluster terms $E^{\rm (1+2BC)}$,
the 3BC terms $E^{\rm (3BC)}$,
and the total ground-state energy $E_{\rm g.s.}$ of $^{16}$O.
The experimental value is taken from Ref.~\cite{Audi93}.
All energies are in MeV.}
\begin{ruledtabular}
    \begin{tabular}{rrrrrr}
          &      &  Nijm I & CD Bonn & N$^{3}$LO & Expt. \\ \hline
 $^{16}$O & $E^{\rm (1+2BC)}$ & $-103.72$ & $-115.58$ & $-105.92$ & \\
          & $E^{\rm (3BC)}$   &   $-4.02$ &   $-3.82$ &  $-13.57$ & \\
& $E_{\rm g.s.}$    & $-107.74$ & $-119.39$ & $-119.48$ & $-127.62$ \\
      &  $BE/A$     &    $6.73$ &    $7.46$ &    $7.47$ & $ 7.98  $ \\
    \end{tabular}
\end{ruledtabular}
\end{table}

In Table~\ref{tab:o16_egs}, we tabulate the energies of
the one- and two-body cluster terms $E^{\rm (1+2BC)}$,
the 3BC terms $E^{\rm (3BC)}$, and the total ground-state energy
$E_{\rm g.s.}$ of $^{16}$O.
The binding energy per nucleon $BE/A=-E_{\rm g.s.}/A$ is also given.
The values of
$\hbar \Omega=14$ MeV and $\rho _{1}=18$ for Nijm I,
$\hbar \Omega=15$ MeV and $\rho _{1}=14$ for CD Bonn,
and $\hbar \Omega=15$ MeV and $\rho _{1}=20$ for N$^{3}$LO
are shown as the optimal ones.
It is seen that, although the 3BC terms have attractive and sizable
contributions to the ground-state energy, the calculated ground-state
energies are still less bound than the experimental value.
In the present calculation, a genuine {\it NNN} force is not
taken into account.
The inclusion of the {\it NNN} force could compensate for
the discrepancies between the theoretical and experimental values,
which has been shown in the microscopic studies of light
nuclei~\cite{Pieper01,Navratil07}.
Note, however, that the energies of $93.6$ \% to the experimental value
are attained from only the NN force for the CD-Bonn and
the N$^{3}$LO potentials, though $84.4$ \% for the Nijm-I interaction.

\begin{figure}[t]
\includegraphics[width=.360\textheight]{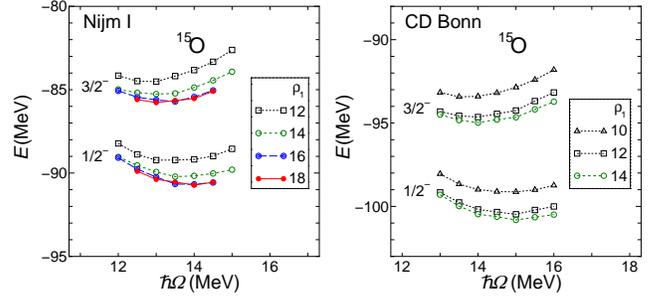}
\caption{\label{fig:o15be} (Color online) The $\hbar \Omega$ and
$\rho_{1}$ dependences of the calculated energies $E(=-BE)$
of the lowest $1/2^{-}$ and $3/2^{-}$ states of the spin-orbit doublet
in $^{15}$O.}
\end{figure}

\begin{table}[b]
\caption{\label{tab:o15be} The calculated energies of the one- and
two-body cluster terms $E^{\rm (1+2BC)}$, the 3BC terms $E^{\rm (3BC)}$,
and the total energy $E(=-BE)$ of the lowest $1/2^{-}$ and $3/2^{-}$
states of the spin-orbit doublets in $^{15}$O and $^{15}$N.
The quantity $E_{\rm s.o.}$ is the spin-orbit splitting energy
including the 3BC effects.
The energy difference $E_{\rm diff}$ of the ground-state energies
between $^{15}$O and $^{15}$N is also given.
All energies are in MeV.}
\begin{ruledtabular}
    \begin{tabular}{rcrrrr}
    & $J^{\pi}$ &      & Nijm I & CD Bonn  & Expt. \\ \hline
$^{15}$O & $3/2^{-}$ & $E^{\rm (1+2BC)}$ & $-81.17$ & $-90.38$ & \\
    &                & $E^{\rm (3BC)}$   &  $-4.61$ &  $-4.59$ & \\
    &                & $E$ & $-85.77$    & $-94.97$ & $-105.78$  \\
         & $1/2^{-}$ & $E^{\rm (1+2BC)}$ & $-85.71$ & $-96.24$ & \\
    & & $E^{\rm (3BC)}$                  &  $-5.01$ & $-4.58$  & \\
    & & $E$                   & $-90.72$ & $-100.81$ & $-111.96$ \\
    & & $E_{\rm s.o.}$        & $4.95$ & $5.84$ & $6.18$  \\ \hline
$^{15}$N & $3/2^{-}$ & $E^{\rm (1+2BC)}$ & $-84.58$ & $-94.00$ & \\
    & & $E^{\rm (3BC)}$                  & $-4.59$  &  $-4.58$ & \\
    & & $E$                    & $-89.17$ & $-98.58$ & $-109.17$ \\
        & $1/2^{-}$  & $E^{\rm (1+2BC)}$ & $-89.14$ & $-99.87$ & \\
    & & $E^{\rm (3BC)}$                    & $-4.99$ & $-4.55$ & \\
    & & $E$                   & $-94.13$ & $-104.42$ & $-115.49$ \\
    & & $E_{\rm s.o.}$         & $4.96$ & $5.84$ & $6.32$ \\ \hline
    & & $E_{\rm diff}$                 & $3.41$ & $3.60$ & $3.54$\\
    \end{tabular}
\end{ruledtabular}
\end{table}

The 3BC effect for N$^{3}$LO is significantly larger than the ones
for Nijm I and CD Bonn.
A similar tendency is seen in the recent $\Lambda$CCSD(T) computation
including triples corrections~\cite{Hagen09}.
We have found that the large 3BC contribution
also applies to the other nuclei in the present study.
Owing to this property, we have not yet obtained the converged results
for the other nuclei.
For this reason, we do not show the other results for N$^{3}$LO.
One may also notice that the result of $E^{\rm (1+2BC)}$ for
N$^{3}$LO shows a large difference of about $4$ MeV from that
in our previous study~\cite{Fujii04}.
This is due to strong dependences of $E^{\rm (1+2BC)}$ and
$E^{\rm (3BC)}$ on $\hbar \Omega$ and $\rho _{1}$ for
N$^{3}$LO.

In Fig.~\ref{fig:o15be}, we illustrate the $\hbar \Omega$ and
$\rho _{1}$ dependences of the total energy $E(=-BE)$ including
the 3BC effects of the lowest $1/2^{-}$ and $3/2^{-}$ states
of the spin-orbit doublet in $^{15}$O.
These states are representative single-hole states of neutron
in $^{15}$O.
For Nijm I, we take the values of $\hbar \Omega=14$ and $13$ MeV
at $\rho _{1}=18$ for the $1/2^{-}$ and $3/2^{-}$ states,
respectively, as the optimal ones, and for CD Bonn,
$\hbar \Omega=15$ and $14$ MeV at $\rho _{1}=14$.
These optimal values are tabulated in Table~\ref{tab:o15be}.
The results for the proton-hole states in $^{15}$N are also given.
The convergence properties for $^{15}$N are similar to the case
of $^{15}$O.

\begin{figure}[t]
\includegraphics[width=.360\textheight]{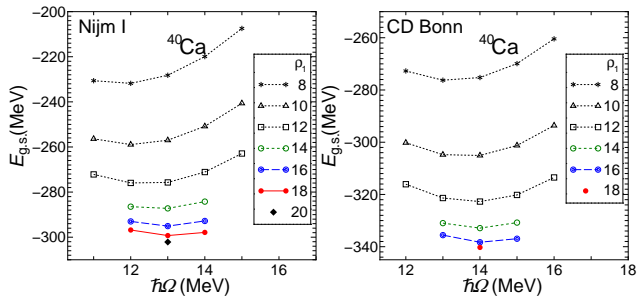}
\caption{\label{fig:ca40egs} (Color online) Same as
Fig.~\ref{fig:o16egs}, but for $^{40}$Ca.}
\end{figure}

\begin{table}[b]
\caption{\label{tab:Ca40_egs} Same as Table~\ref{tab:o16_egs},
but for $^{40}$Ca.}
\begin{ruledtabular}
    \begin{tabular}{rrrrr}
           &      &    Nijm I  & CD Bonn   & Expt. \\ \hline
 $^{40}$Ca & $E^{\rm (1+2BC)}$ & $-296.29$ & $-334.34$ & \\
           & $E^{\rm (3BC)}$   &   $-5.83$ &   $-5.92$ & \\
  & $E_{\rm g.s.}$ & $-302.12$ & $-340.27$ & $-342.05$ \\
           &  $BE/A$       & $7.55$ & $8.51$ & $ 8.55  $ \\
    \end{tabular}
\end{ruledtabular}
\end{table}

The microscopic description of the spin-orbit splitting in nuclei
is a long-standing problem.
In Table~\ref{tab:o15be}, the spin-orbit splitting is denoted by
$E_{\rm s.o.}$ which is the difference of the
binding energies between the $1/2^{-}$ and $3/2^{-}$ states.
Our results
show smaller splitting energies than the experimental
values for these hole states,
which does not contradict previous studies~\cite{Ando81, Pieper93}.
We should note, however, that the magnitude of the lack of the
splitting energy depends considerably on the interactions employed.
For CD Bonn, the differences are only $0.34$ and $0.48$ MeV
for $^{15}$O and $^{15}$N, respectively.
The calculation including the {\it NNN} force could give
a better result as shown in Refs.~\cite{Ando81, Pieper93}.

The energy difference of the ground states between $^{15}$O and
$^{15}$N is denoted by $E_{\rm diff}$ in Table~\ref{tab:o15be}.
Our results are in good agreement with the experiment.
Similar tendency has been found in the case of $^{3}$He and $^{3}$H
in our previous work~\cite{Fujii04}.
Since we include the Coulomb force, the small differences
between the results and experiment may be attributed to
the effects of the charge-independence breaking of the
original {\it NN} forces.

In Fig.~\ref{fig:ca40egs}, 
we show the $\hbar \Omega$ and $\rho_{1}$ dependences of the calculated
ground-state energies including the 3BC effects of $^{40}$Ca.
We take the values of
$\hbar \Omega =13$ MeV and $\rho _{1}=20$ for Nijm I,
and $\hbar \Omega =14$ MeV and $\rho _{1}=18$ for CD Bonn
as the optimal values.
Since we handle a heavier system $^{40}$Ca than $^{16}$O,
we need a larger model space to obtain the converged results.
The optimal values are given in Table~\ref{tab:Ca40_egs}.
The results of $E_{\rm g.s.}$ are less attractive than
the experimental value similarly to the case of $^{16}$O.
However, for CD Bonn, the calculation attains $99.5$ \%
of the experimental energy, and the difference
between the result and experiment is only $1.78$ MeV.
This difference is much smaller than that for $^{16}$O despite
the fact that the absolute value of the ground-state energy of
$^{40}$Ca is much larger than that of $^{16}$O.
This fact suggests that the {\it NNN} force plays a complicated role
in the inner (dense) and outer (thin) regions of the nuclei.

\begin{figure}[t]
\includegraphics[width=.360\textheight]{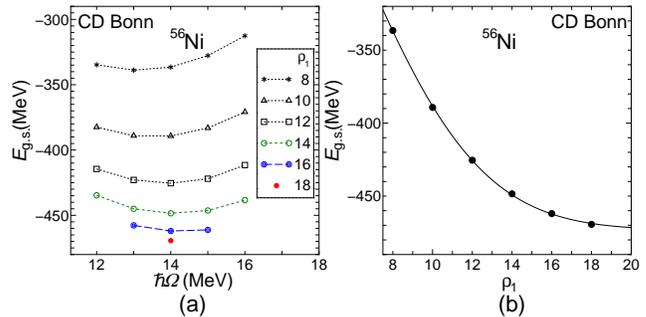}
\caption{\label{fig:ni56egs} (a) (Color online) Same as
Fig.~\ref{fig:o16egs}, but for $^{56}$Ni.
(b) The extrapolation curve of
$E_{\rm g.s.}$ of $^{56}$Ni. See text for details.}
\end{figure}

\begin{table}[b]
\caption{\label{tab:Ni56_egs}  Same as Table~\ref{tab:o16_egs},
but for $^{56}$Ni.
The results only for the CD-Bonn potential are shown.
The extrapolated values for $\rho_{1}\rightarrow \infty$
are also given.}
\begin{ruledtabular}
    \begin{tabular}{rrrrr}
  & & $\rho_{1}=18$ & $\rho_{1}\rightarrow \infty$ & Expt. \\ \hline
  $^{56}$Ni        & $E^{\rm (1+2BC)}$ & $-452.73$ & $-456.64$ & \\
                     & $E^{\rm (3BC)}$ &  $-16.62$ &  $-16.53$ & \\
         &  $E_{\rm g.s.}$ & $-469.35$ & $-473.17$ & $-483.99$ \\
                     & $BE/A$ & $8.38$ &    $8.45$ &    $8.64$ \\
    \end{tabular}
\end{ruledtabular}
\end{table}

In Fig.~\ref{fig:ni56egs}(a), the $\hbar \Omega$ and $\rho_{1}$
dependences of the calculated ground-state energies including
the 3BC effects of $^{56}$Ni are illustrated for CD Bonn.
A converging result is seen at $\hbar \Omega=14$ MeV and
$\rho _{1}=18$.
However, the energy difference between the two values for
$\rho _{1}=16$ and $18$ at the energy minima amounts to $7.38$ MeV
which is somewhat large compared to the case of $^{40}$Ca
where the difference is $1.97$ MeV.
The results for the CD-Bonn potential show a regular pattern of
convergence
(in contrast to the results for N$^{3}$LO in Fig.~\ref{fig:o16egs}).
In order to estimate the remaining effect of the larger model space,
we perform an extrapolation,
as given in Refs.~\cite{Suzuki87, Suzuki94},
using the following formula:
$E_{\rm g.s.}(\rho _{1})=E_{\infty}+Ce^{-\gamma \rho _{1}^{2}}$,
where $E_{\infty}$, $C$, and $\gamma$ are the coefficients determined
in the least-squares fitting procedure.
We have found that the data points for $\hbar \Omega =14$ MeV from
$\rho _{1}=8$ to $18$ are well fitted with this formula.
The curve given by the formula is shown in Fig.~\ref{fig:ni56egs}(b).
The optimal values of the coefficients are $E_{\infty}=-473.17$ MeV,
$C=-316.06$ MeV, and $\gamma=1.3148\times 10^{-2}$.
Therefore, the extrapolated ground-state energy for
$\rho _{1} \rightarrow \infty$ becomes
$E_{\rm g.s.}(\rho _{1}\rightarrow \infty)=E_{\infty}=-473.17$ MeV.
The difference between the extrapolated value and the result for
$\rho _{1}=18$ is $3.82$ MeV, and thus the result for $\rho _{1}=18$
is considered to be an almost converged value.
The extrapolated value reproduces $97.8$ \% of the experimental
ground-state energy.

In Table~\ref{tab:Ni56_egs}, the optimal value of
$\hbar \Omega =14$ MeV and $\rho _{1}=18$ and the extrapolated one
for $^{56}$Ni using the CD-Bonn potential are listed.
The extrapolation for $E^{\rm (1+2BC)}$ in the same manner has been
performed, and its result is also given.
It is seen that the 3BC effect of $^{56}$Ni is considerably larger than
that of $^{40}$Ca.
This may reflect the difference of the shell closure, namely,
$0f_{7/2}$ sub-shell
closed for $^{56}$Ni and $1s0d$ major-shell closed for $^{40}$Ca.

In summary, we have applied the UMOA to the ground states of
the closed-shell nuclei $^{16}$O, $^{40}$Ca, and $^{56}$Ni,
and the single-hole states in $^{15}$O and $^{15}$N.
The $pf$-shell nucleus $^{56}$Ni is the heaviest one for which
this kind of {\it ab initio} calculation has been performed.
The binding energies including the 3BC effects have been obtained
using the Nijm-I, the CD-Bonn, and the chiral N$^{3}$LO {\it NN}
interactions.
We have found that the chiral N$^{3}$LO interaction gives rather
large 3BC contribution to the ground-state energy of $^{16}$O
compared to the other two forces.
All results lack the binding energies
in reproducing the experimental data.
However, the magnitude of the missing energy depends considerably on
the interactions employed.
The CD-Bonn potential gives quite good results close to
the experimental values for all nuclei in the present work.
The inclusion of the {\it NNN} force is expected to make attractive
contributions and compensate for the remaining discrepancies
between the results and experiments.

One may compare the present results with recent microscopic
calculations~\cite{Hagen08,Hagen09,Coraggio06,Barbieri06,Gour06}
using realistic {\it NN} forces including the ones used here.
Although the present and recent methods give similar results,
there still remain some discrepancies.
It is an important problem to clarify the origins of the discrepancies
in order to develop the microscopic many-body methods.

For a deeper understanding of nuclei, the use of more fundamental
forces is of great interest.
Recently, a novel {\it NN} force from a lattice QCD calculation has
been reported, and a more elaborate work of nuclear force is
in progress~\cite{Ishii07}.
We will pursue studies using forces based upon the lattice QCD.

\begin{acknowledgments}

This work was supported
 by a Grant-in-Aid for Young Scientists (B)
(No. 18740133) from JSPS, the JSPS Core-to-Core Program EFES,
and Grants-in-Aid for Scientific Research on Innovative Areas
(Nos. 20105001 and 20105003) from MEXT.

\end{acknowledgments}

\end{document}